\documentclass[journal]{IEEEtran}

\usepackage{graphicx}
\usepackage{tablefootnote}
\usepackage{multirow}
\usepackage{bchart}

\begin{document}
\title{Personalized Chatbot Trustworthiness Ratings}
%\author{Biplav Srivastava, Francesca Rossi,  Sheema Usmani and Mariana Bernagozzi}
%\author{Paper ID: 50}
%\author{Paper ID: 2765}
%\author{Paper ID: 5317}
%\author{All Contributors - anonymized for submission}

%\author{Biplav Srivastava$^1$ and Francesca Rossi$^2$ and Sheema Usmani$^1$ and Mariana Bernagozzi$^1$\\
%$^1$ IBM Chief Analytics Office, $^2$IBM T. J. Watson Research Center}

\author{Biplav~Srivastava$^1$,~\IEEEmembership{Senior Member,~IEEE,}
        Francesca~Rossi$^2$, Sheema~Usmani$^3$ and Mariana~Bernagozzi$^3$
        %and~Jane~Doe,~\IEEEmembership{Life~Fellow,~IEEE}% <-this % stops a space
\thanks{$^1$AI Institute, University of South Carolina, Most of the work was done while Biplav Srivastava was at IBM. $^2$IBM T. J. Watson Research Center, $^3$IBM Global Business Services. }
%\thanks{M. Shell was with the Department
%of Electrical and Computer Engineering, Georgia Institute of Technology, Atlanta,
\thanks{Manuscript submitted on January 23, 2020.}}

\markboth{2020-01-0004-OA10-TTS - Personalized Chatbot Trustworthiness Ratings, Jan~2020}
{Shell \MakeLowercase{\textit{et al.}}: IEEE Transactions on Technology and Society}


\maketitle

% -------------------------------------------
\begin{abstract}
%\begin{quote}
%Conversation agents, commonly referred to as chatbots, are increasingly deployed in many domains
%to allow people to have a natural interaction with machines while trying to solve a specific problem. Given their widespread use, it is important to provide users with methods and tools to increase awareness of various properties, including non-functional properties that users may however consider important in order to trust a chatbot.  
%For example, users may want to use chatbots that are not biased, that do not use abusive language, that do not leak information to other users, and that respond in a style which is appropriate for the user's cognitive level. 
%However, as the field matures and response generation moves from static to dynamic utterances which are sensitive to user's unique context and dynamic content, it is difficult to ensure such trust-related issues, both to developers and to users. 

In this paper, we address a setting where a conversation agent, also know as a chatbot, cannot be modified and its training data cannot be accessed, and yet a neutral party wants to assess and communicate its trustworthiness to a user in a way that is tailored to the user's priorities over the various trust issues (such as bias, abusive language, information leakage, or inappropriate communication complexity). 
%of in the context of a user's priorities over the issues. 
Such a rating can help users choose among alternative chatbots, developers test their systems, business leaders price their offerings, and regulators set policies. We describe a chatbot rating methodology that relies on separate rating modules for each trust issue, and on users' priority orderings among the relevant trust issues, to generate an aggregate personalized rating for the trustworthiness of a chatbot.
%for a certain user profile.  
The method is independent of the specific trust issues and is parametric to the aggregation procedure, thereby allowing for seamless generalization. We illustrate its general use, integrate it with a live chatbot, and evaluate it on four dialog datasets and representative user profiles, validated with a user survey.

\end{abstract}

% -------------------------------------------


% -------------------------------------------

\IEEEpeerreviewmaketitle

% -------------------------------------------

% ----------------------------------------
% ----------------------------------------
\section{Introduction}

% -----------------------------------------

\begin{figure}
\centering
  \includegraphics[width=0.4\textwidth]{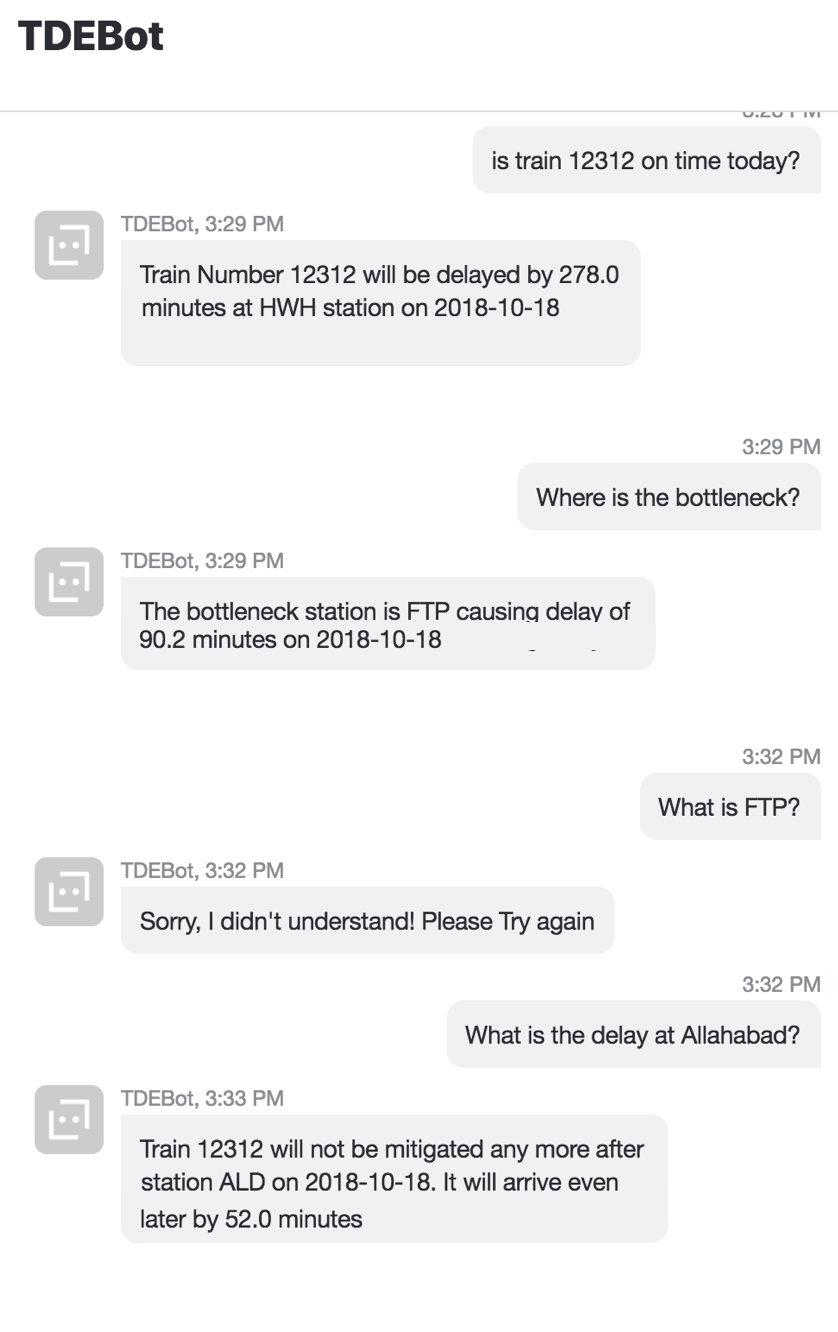}
    \caption{Sample interaction of a user with a train assistance chatbot for Indian railways \cite{train-chatbot}. Trust issues: chatbot becoming abusive, leaking user travel information to other users and developers, having incomprehensible conversational style.}
 \label{fig:train-bot}
\end{figure}
% -----------------------------------------

Conversation is a hallmark of intelligence and a major way in which humans communicate. 
This is why conversation agents, commonly referred to as chatbots, are increasingly deployed in many domains
to allow people to have a natural interaction with machines while trying to solve a specific problem. 
Businesses wanting to build AI-based systems to increase productivity and improve customer experience are therefore interested in building various forms of automated conversation systems, dialog systems, or chatbots. There are many platforms available to create chatbots \cite{chatbot-survey-accenture,dialog-intro}.  

%A recent paper reviews the state-of-art and looks at requirements and design options to make them customizable for end-users as their own personal bot \cite{chatbot-arch-icse18}.

% Most common types of chatbots deal with a single user at a time and conduct informal conversation, answer the user's questions or provide recommendations in a given domain. 
% A key problem in building chatbots is that of dialog management, i.e., creating dialog responses to user's utterances. The two key    approaches for building them are learning policies over dialog examples \cite{young2013pomdp} or  reasoning on its abstract representations \cite{minim-dialog}.
% A recent survey looks at advances in (deep) learning methods in building systems for general  and task-oriented conversations  \cite{chatbot-survey}. 

\begin{table*}
%\vspace*{-7mm}
{\small
\begin{center}
%\begin{tabular}{|c|l|c|l|c|}\hline
%\begin{tabu} to 0.8\textwidth{|X[l]|X[l]|X[l]|X[l]|X[l]|}\hline
%\begin{minipage}{\textwidth}
\begin{tabular}{|c|l|c|l|c|}\hline
{\bf Bias} &{High }&{Score}&{Low }&{Score}
  \\\hline\hline
Ubuntu & Mint seems better & 1 
  & no i just configured it	& 0\\\hline
Insurance  &  Which Company Has The Best  & 1	
   & What Does Split Limits Mean In 	& 0 \\
   &   Retirement Plan? & 
   & Auto Insurance?	& \\\hline
HR & Works pretty well needs some work with,  	
  & 0.69 & 
  I need to search the intranet to find an answer 	
  & 0.01\\
  & better answers fantastic for simple questions 	
  &  & to your question. & \\
  & and quick info	
  &  &  & \\\hline
Restaurant & i hope prefer expensive restaurant	& 0.72 & You are looking for a restaurant is that right?	& 0\\\hline \hline
% ---
{\bf Abuse} &{High }&{Score}&{Low }&{Score}
  \\\hline\hline
Ubuntu & bazang is a f*g\tablefootnote{Used "*" to replace letter "a" in the original text.} & 1 & and then take a look at the iptables?	& 0\\\hline
%Insurance  & no you not can get Life Insurance on a fetus the 	child have be & 1	& Can A Life Insurance Claim Be Denied?	& 0\\
Insurance  & no you not can get Life Insurance on a fetus & 1	& Can A Life Insurance Claim Be Denied?	& 0\\
  & the child have be born in most case at least & 	& 	& \\
  & 14 day old old before you can consider ...	
  &	& & \\\hline
HR& Hi Chip, hoe do I setup Lotus notes?    & 0.5 &I don't know the answer to your question. 	& 0\\
  &  & & Let me try to find it on the intranet for you. & \\\hline
Restaurant & -	& - & pizza hut cherry hinton is a great restaurant	& 0\\\hline \hline
% ---
{\bf Complexity} &{High }&{Score}&{Low }&{Score}
  \\\hline\hline
Ubuntu & sudo adduser user group & 1 & that's my impressions	& 0.25\\\hline
Insurance  &will homeowners insurance cover flooring?	& 1	& what are some examples of annuities?	& 0.5\\\hline
HR& are company email addresses case sensitive?	& 0.92 &where am i?	& 0.33\\\hline
Restaurant  & the lucky star serves Chinese food	& 0.94 & coke it is	& 0.33\\\hline
\end{tabular}
\vspace*{2mm}
\caption{Examples of utterances with high and low scores by issues.}
\label{tab:eg}
%\end{minipage}
\end{center}
}

\end{table*}

% -------------

However, chatbots can be fraught with ethical issues. An extreme %and anecdotal 
example is the Tay \cite{tay,tay-comments} Twitter chatbot, released by Microsoft in 2016, that was designed to engage with people on open topics and learn from feedback, but ended up getting manipulated by users to exhibit unacceptable behavior via its extreme responses. During the COVID19 pandemic, some people have interacted with chatbots to overcome loneliness and found solace but this has raised various issues\cite{covid-bot}. Another example is shown in Figure~\ref{fig:train-bot}, where a chatbot to answer train delay information may raise unexpected trust issues \cite{train-chatbot}.  It can become abusive, leak user travel information to other users and developers and have incomprehensible conversational style (see more examples in Illustration section later). 
%related to using a terminology that is not easily understood by some classes of users.
More generally, several potential ethical issues in dialogue systems built using learning methods can be identified, such as 
showing implicit biases from data, being prone to adversarial examples, being vulnerable to privacy violation, the need to maintain safety
of people, and concerns about explainability and reproducibility of responses \cite{ethical-dialog}.

Given the widespread and increasing use of chatbots, it is important to address such ethical concerns and provide users with methods and tools to increase awareness of various properties, including non-functional properties that users may however consider important in order to trust a chatbot. For example, users may want to use chatbots that are not biased, that do not use abusive language, that do not leak information to other users, and that respond in a style which is appropriate for the user's cognitive level. 

% ----------------------------------------
\subsection{Interested Parties and Their Concerns}
%\noindent{\bf Interested Parties and Their Concerns}
After a chatbot is built, it is usually deployed as a shared service available to users via Internet-connected devices (such as computers or mobiles) or embodied systems (such as robots, Amazon Alexa, or Google Home). In the process of interaction, data is passed between the user and the chatbot service provider.
Many interested parties could be concerned about the chatbot's behavior. They include at least the following ones:  

%\vspace{-.5cm}
\noindent {\bf Users}: Users are 
%the ones interacting with the chatbot. They are 
 concerned with the value they derive by interacting with the chatbot and expect it to follow social and business norms similar to those they expect from human assistants. One example is that the user's information, whether sensitive or otherwise, be not revealed to other users without their permission. The repercussion of such a breach is not only loss of trust but can also be illegal, depending on the type of user (such as child, adult, patient, celebrity, government official) and context of usage. 
This is further complicated by the fact that, over time, a chatbot may get personalized to a user's need but the person may not want to share her personalized information with the developers.
%Moreover, shared information may spread when other users interact over time \cite{info-spread}.
Another user's concern is the possible use of abusive language by the chatbot. 
Yet another issue could be the use of language that the user does not understand, because too complex or not appropriate for his/her knowledge of the subject.
% A chatbot may be deployed as a shared service but made available to users as-is or embodied in a internet-connected physical device like robot, Amazon Alexa, Google Home. 

%\vspace{.5cm}
\noindent {\bf Developers}: Developers want the chatbot to perform as per design and may worry that it could say something which gives an unintended perception to human users. Examples of developers' concerns are related to bias behavior (that is, the chatbot should not be prone to erratic response in the presence of protected variables like gender or race) and language usage (that is, the chatbot should not respond with  hateful or abusive language).

%\vspace{.5cm}
\noindent {\bf Data providers}: They provide data that is used by chatbots' developers for the training phase, ranging from the domain of discourse (e.g., financial data), encyclopedic information (e.g., Wikipedia), to language (e.g., Word embeddings). Data providers want to make sure their data is of the best quality feasible.

%\vspace{-3.5mm}
\subsection{Approach and Contributions}

In this paper, we address the setting where a chatbot cannot be modified and its training data cannot be accessed, and a neutral party wants to assess and communicate the trustworthiness of chatbots in the context
of a user's priorities over the issues. Such a rating can help users choose among alternative chatbots, developers test their systems, business leaders price their offerings, and regulators set policies. We envision a personalized rating methodology for chatbots that relies on separate rating modules for each issue, and users' detected priority orderings among the issues, to generate an aggregate personalized rating for the trustworthiness of a chatbot for a certain user profile.  
We focus on 4 issues: %(1) 
Fairness and bias ($B$), %(2) 
Information leakage ($IL$),
Hate and Abusive language ($AL$), and Conversation Complexity ($CC$). Table~\ref{tab:eg} illustrates these issues on some dialog datasets.
However, our framework and methodology is general and can be extended to other issues. % as well.

%\noindent{\bf Our contributions}

We make the following  contributions:
%\begin{itemize}
%\item 
(a) introduce the notion of a contextualized rating of the trustworthiness of a chatbot; %with well-defined semantics which consists of an expandable set of issues;
%\item 
(b) propose a method to compute the rating by using relative importance rankings over issues, provided by users;
%\item 
(c) present an architecture to implement the rating approach as a service;
%\item 
(d) integrate the method with a live chatbot;
(e) evaluate our approach on four dialog datasets 
and representative user profiles which we validate in a user survey.
%\end{itemize}

% The rest of the paper is organized as follows: we
% start with a background on chatbots
% and describe some possible trust issues in using such systems.
% We then present our setting and rating method, and illustrate it in the context of two chatbots: a chit-chat one and a task-oriented one. We then describe an implementation and present results on four dialog datasets and one live chatbot. Finally, we conclude mentioning some related work. 

% ----------------------------------------

\section{Related Work}

The brittleness of machine learning models is well known,
and chatbots represent a specific usage of such models. 
For NLP models, \cite{sear-acl18} presented a method to
generate syntactically different but semantically equivalent test cases that can 
flip the prediction of the model. 
This can be a cause of concern to application developers for usability reasons, and addressing it can prevent the system from being exploited by an adversary. In the context of chatbots, the concern is that its output can be manipulated by changing the input and this is what the sensitivity test can detect.
%As noted before, t

The authors in \cite{ethical-dialog} systematically  survey a number of potential ethical issues in dialogue systems built using learning methods. However, they do not consider a method to communicate a trustworthiness rating based on the analysis of such issues. 

There is a rich body of work on studying issues influencing online services and AI methods. 
In information spreading, the seminal work is described in \cite{info-spread} where the authors looked at the spread of information in social networks and how to maximize it %s spread 
by engaging the effective influencer nodes. 
%The basic information diffusion models are: {\em Linear Threshold Model}, which considers influence of neighbors and thresholds to determine which nodes are subsequently influenced (activated from inactive state), and {\em Independent Cascade Model}, which considers a node given a single chance to influence its neighbors independent of its history. These models can be used to design information leak checkers.
In studying abusive language online, the authors in \cite{twitter-curse} explore the prevalence of cursing on Twitter which serves as a platform for utterance and conversation. They found that people curse more online than in physical environment, among same gender, when they are angry or sad, as their activities increase during the day, and when in relaxed or formal environments. But users may not want the chatbots they are interacting with to exhibit the same behavior, especially when the users are children. 

The closest prior work to what we describe in this paper is on rating AI services\cite{rating-aies2018}. There, the authors propose a 2-step bias rating procedure for invocable one-shot AI services (like translation service), as well as a composition method to build sequences of such services. However, that work does not consider: (a) multiple issues and users, (b) personalized rating based on users' ranking of issues, (c) dialog setting of a series of interactions, and not just a single invocation, and (d) conversations leading to completion of tasks.

% There are many definitions of bias. An agent can be considered bias for a number of reasons. Some are:
% \begin{itemize}
%   \item For not using protected variables in a way that 
%     is considered unbiased.
%   \item For using hate speech   
%       \cite{hateoffensive,twitter-curse}.
% \item 
% \end{itemize}

% ----------------------------------------

% ----------------------------------------

% --------------------
\section{Dialog Systems and Response Generation}
%\section{Preliminaries -- Dialog and Question Answering with Learning and Reasoning}

% Chatbots \cite{dialog-intro}, which can engage people in natural dialog conversations, have recently gained more popularity because of numerous platforms to create them quickly for any domain \cite{chatbot-survey-accenture}. Most common types of such AI agents deal with a single user at a time and conduct informal conversation, answer the user's questions, or provide recommendations in a given domain.
% They need to handle uncertainties related to human behavior and natural language, while conducting some form of dialog to achieve their goals. 
%A very significant area for chatbots deployment is customer support in many industries, where they are expected to save over \$8 billion per year by 2022 \cite{chatbot-cc-juniper}. 

% % -----------------------------------------

% \begin{figure*}
%  \centering
%   \includegraphics[width=0.7\textwidth]{figs/chatbot-arch.jpg}
%   \caption{The architecture of a data-driven chatbot.}
%   \label{fig:chatbot-arch}
% \end{figure*}
% % -----------------------------------------

A {\em dialog} is made up of a series of {\em turns}, where each turn is a series of {\em utterances} by one or more participants playing one or more {\em roles}. For example, in the customer support setting,  the roles are the {\em customer} and the {\em support chatbot}.
%Common dialog agents deal with a single user at a time and conduct informal conversation, answer the user's questions or provide recommendations in a given domain.
%dialogs to achieve their goals. 

The core problem in building chatbots is that of dialog management (DM), i.e., creating useful dialog responses to the user's utterances \cite{dialog-intro}.
%The system architecture of a typical data-driven chatbot is shown in Figure~\ref{fig:chatbot-arch}.
There are many approaches to tackle DM in literature, including finite-space, frame-based, inference-based, and statistical learning-based \cite{chatbot-survey-statistical-ml,chatbot-book,minim-dialog,young2013pomdp}, of which  finite-space and frame-based are the most popular ones with mainstream developers. 
In a representative invocation, the user's utterance is analyzed to detect her intent and a 
policy for response is selected. This policy may call for querying a database and the result of query execution is %returned, which is 
then used by the response generator to create a response, usually using some templates. 
The system can dynamically create one or more queries, which involves selecting tables and attributes, filtering values, and testing for conditions,  and assume defaults for missing values. 
It may also decide not to answer a request if it is unsure of the correctness of a query's result.
Note that the DM module may use one or more domain-specific databases %(sources) 
as well as one or more domain-independent sources like language models and word embeddings. The latter has been found to be a possible source of human bias  \cite{lang-bias-nature}.
\section{Trust Issues in Dialog Services} %AI Services}

Trust is a very important factor in AI development, deployment, and usage. 
%If we want AI to be widely adopted, so to get all its positive impact, we need to build a system of trust around it. 
Users and stakeholders should be able to have a justified trust in the AI systems they use, otherwise they will not adopt them in their everyday life. 
%This is especially important in setting where there are high stake decisions to be made, such as healthcare. 
In general, there are various dimensions of trust to be considered, that range from robustness to reliability, and from transparency to explainability and fairness. We focus on a subset of issues whose checkers are available and robust.
% Here, we consider some of these trust dimensions in chatbots, specifically regarding abusive language, bias, user privacy and conversation complexity. These four issues also happen to have already robust checkers. 

%Here we focus on some trust dimensions that are specific to chatbots, and that aim to build trust in the fact that the chatbot does not behave in ways that are not desired, especially regarding abusive language, bias, information protection and privacy, and conversation complexity. 
%Let us define them more in detail:

%\vspace{.5cm}
\noindent {\bf Abusive Language:} An important issue in the usage of a chatbot is the possibility of hate and abusive speech. This can make the chatbot unacceptable or inappropriate to some users, harm people in unintended ways, and expose service providers to unknown risks and costs. There is a growing body of work to detect hate 
speech\cite{hateoffensive} and abusive language\cite{twitter-curse} online using words and phrases which people have annotated. The authors in the former paper define hate speech  as {\em language that is used to expresses hatred towards a targeted group or is intended to be derogatory, to humiliate, or to insult the members of the group}. Their checker, which we use in our work, has a logistic regression with L2 regularization to achieve automatic detection of hate speech and offensive language. 

%\vspace{.5cm}
\noindent {\bf Bias:} Another issue with AI services is the presence of bias. Bias can result in an unfair treatment for certain groups compared to others, which is undesirable and often illegal.
%this is not desired and often not legal.
There are many definitions of fairness, each one suitable for certain scenarios. \cite{fairness-github} introduces 
notations and a few definitions of fairness while 
\cite{ethical-dialog} discusses bias in dialog systems.
%For example, in \cite{cf-fairness}
%counterfactual fairness is defined, which captures the notion
%that, given two different groups of a protected variable (such as gender or race), a decision is fair towards an individual of a group if it is the same in (a) the actual world and in (b) a counterfactual world, where the 
%individual belongs to the other group.
%In \cite{fair-aware}, instead, the authors describe some measures of statistical fairness. 
Also, \cite{themis} introduces a software testing framework for bias where they use the notions of group and causal bias, and in \cite{aif360} the authors describe a tool to explore fairness using a sample of data, methods, and criteria. In this paper, we use the bias checker implemented by
 \cite{Hutto2015ComputationallyDA}.
%In \cite{rating-aies2018}, the authors propose a testing methodology to rate bias in AI services and to derive bias information by composing several AI modules, showing results on text-based AI services, such as automated language translators.

%\vspace{.5cm}
\noindent {\bf Information Leakage:} This issue involves ensuring that  information given by users to a chatbot is not released, even inadvertently, to other users of the same chatbot or the same platform.
This is further complicated by the fact that over time, a chatbot may get personalized to a user's need but the person may not want to share her personalized information with the developers.
Moreover, shared information may spread when other users interact \cite{info-spread} 
over time. We use the framework discussed in \cite{ethical-dialog} to check this issue.

%\vspace{.5cm}
\noindent {\bf Conversation Style and Complexity:} This issue has to do with making sure that AI services interact with users in the most useful and seamless way. If a chatbot responds to user's questions with a terminology that the user is not familiar with, she will not get the required information and will not be able to solve the problem at hand. 
%So it is important that a dialog system uses a language that the user can understand and find useful for the purposes of the interaction. 

In \cite{dialog-complexity}, the authors propose 
a measure of \textit{dialog complexity} to characterize 
how participants in a conversation use words to
express themselves (utterances), switch roles
and talk iteratively to create turns, and  span the
dialog. They measure the complexity of service dialogs at the levels of utterances, turns and overall dialogs. The method takes into consideration the concentration of domain-specific terms as a reflection of user request specificity, as well as the structure of the dialogs as a reflection of user's demand for (service) actions. We use their checker for implementation.

%In \cite{dialog-complexity}, the authors presented a  data-driven method to assess the complexity of a conversation at different levels of granularity (i.e., utterance, turn and dialog).
%, and use it in a customer-support application. 
%They use a corpus of dialogs to discover domain-specific terms and use it along with dialog length to calculate complexity.

% --------------------

\section{Our Setting and the Rating Method}

% -----------------------------------------

\begin{figure*}
 \centering
   \includegraphics[width=0.9\textwidth]{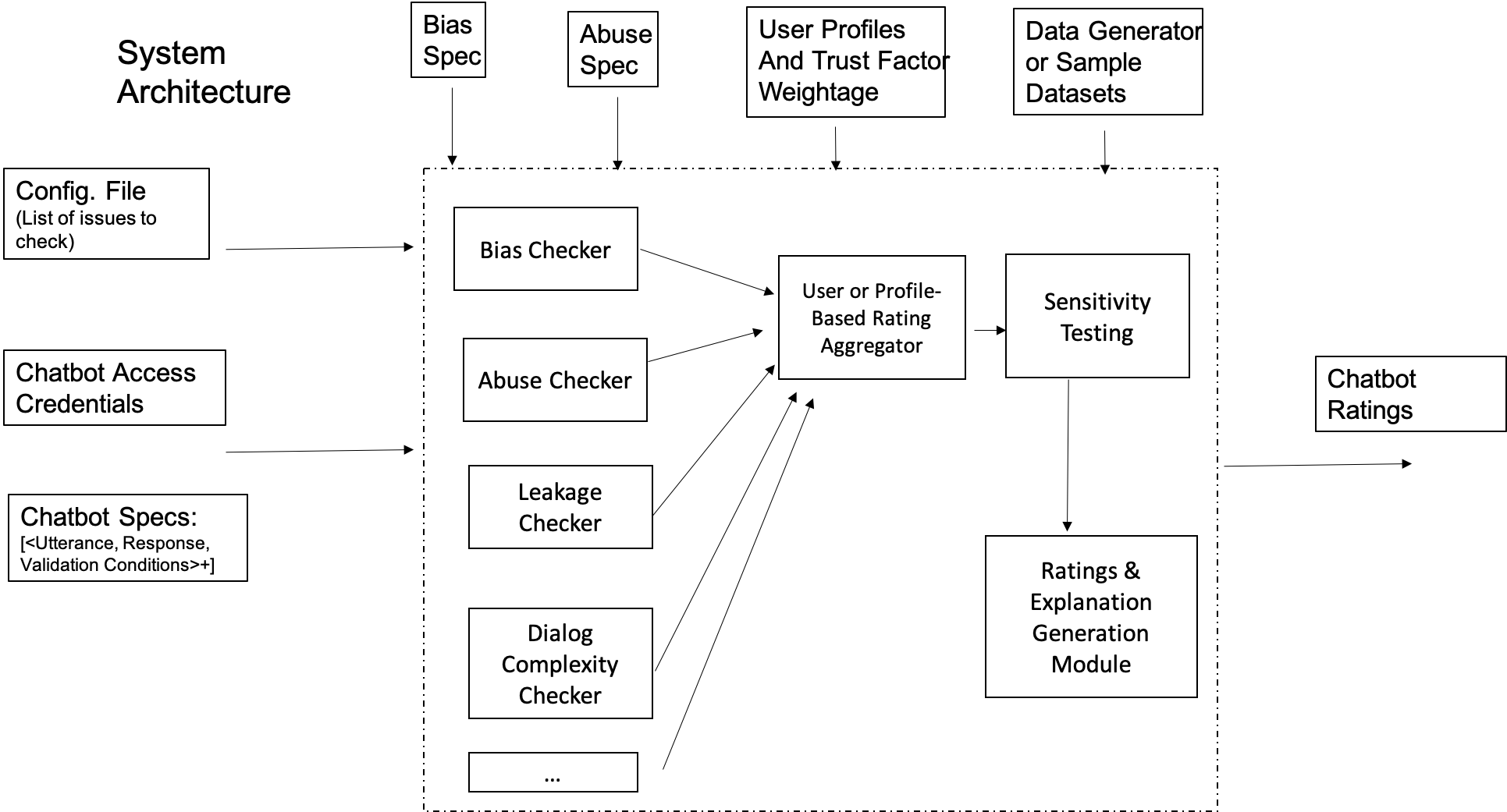}
  \caption{Building blocks of our chatbot rating system.}
  \label{fig:sol-arch}
\end{figure*}
% -----------------------------------------

In this section, we describe our approach to chatbot trust rating.
In particular, we describe the considered setting, the system architecture, and the five steps of our approach. We also highlight the four choice dimensions that need to be instantiated when creating one instance of the general approach.

We consider a setting where a dialog system 
is rated for its behavior with a list of {\em configurable}
%, especially regarding
%how it behaves in terms of a list of 
$k$ issues, such as bias (B), abusive language (AL), conversation complexity (CC), and information leakage (IL). 
The system
is conceptually illustrated in Figure~\ref{fig:sol-arch}.
Its inputs are the issues to be considered, the details of the chatbot to be rated, a user profile, and the
(query) datasets to use for the test,
and its output is a rating
for the chatbot, conveying its level of trustworthiness for a specific user or user's profile.
%and sensitivity to model, data and user.
We will now describe the modules and steps of the proposed rating methodology:

\noindent{\bf 1. Get individual ratings from issue checkers.}
We assume that we have  one checker available for each issue, which can rate the behavior of the dialog system on that issue on a 3-level trust risk scale: [Low, Medium, High] (High meaning that the chatbot is not behaving well regarding that issue). 
For issues with raw scores in a continuous [0-1] range, we bin
them into the 3-level scale.
For $k$ issues, we therefore get a list with $k$ elements
in [Low, Medium, High].
%So what we start with is a list of $k$ elements (where $k$ is the number of issues) from the Low/Medium/High scale. 

\noindent{\bf 2. Elicit/learn users' importance orders.}
The second step involves the aggregation of the elements 
of such 
a list into a single element from that same scale,
in order to get a single rating for the trustworthiness of the chatbot.
We propose to do that by asking  users about the relative importance of the various issues. 

Preference elicitation can be done by asking users about the relative importance of the various issues ({\em individual-level modeling}) or capturing preferences of people as groups and validating them via surveys({\em profile-level modeling}). 
%This will allow us to perform the aggregation in a personalized way. In fact, 
Individual-level models are accurate but hard to build and manage (due to privacy considerations) and generalize. Profile-level models are representative of people who identify with them and easier to implement. Regardless of the granularity, we believe that trustworthiness is not an absolute property, but rather relative to each user (or user profile) of a chatbot.
%that it depends on who is going to use the chatbot. 
%What can be trustworthy, and thus adopted and used with positive impact, for a user may be unacceptable for another one. 
We build profile-level user models, validate these models using a survey, and test them on dialog datasets. New profiles can be added and existing ones updated based on survey responses to capture the preferences of the user base. 

% So we elicit/learn from each user
% her ranked preference order over the $k$ issues, which should represent the levels of importance of the various issues for that user. For example, if a user is very worried about bias and abusive language, but not so much about information leakage and conversation complexity, she could provide the following ranking: B, then AL, then CC, then IL, where issues listed earlier are interpreted as being more important to the user. 
% Notice that we ask users for a qualitative order (and not 
% quantitative values), since we believe that such an elicitation/learning task is easier for the user and faster to work with. Such preference elicitation can be done at the level of each user or at that of a user profile, which 
% %is a set of well-defined characteristics that users are assigned to and 
% defines a class of users which shares the same ranking of issues.

%Notice that we are not proposing to collect quantitative information from the users, that is, how much a user considers an issues more important than another one, but just a qualitative order. While this means we have to work with less information, we believe that this is compensated by the fact that the elicitation/learning task is easier for the user and faster to perform. 

%Notice that this preference elicitation can be done at the level of a single user or at that of a user profile, that is, a set of well-defined characteristics that users are assigned to.

\noindent{\bf 3. If elicitation is done at the user level, aggregate importance orders from similar users.}
%In step 1 we got a rating for each issue, while in step 2 we got a preference order for all issues from each user. 
We would like a single preference order, not many, so that we can then combine the rating levels of the various issues according to this single order and therefore get to a single trustworthiness rating for the chatbot.
However, we do not want to aggregate over all users, but only over similar users, according to some notion of similarity. In this way, the rating will be personalized for each user group, which includes users that are similar to each other.
%Once we have a ranked order from each user (or profile), we propose to aggregate these orders by provided by several users if we cast the users within the same class, according to some notion of similarity. 
The task is therefore to aggregate several ranked orders. To do this, one can use a voting rule, such as Plurality, Borda, Approval, Copeland, etc., as defined in voting theory \cite{vote-counting-survey}. 
We worked with user profiles, and hence, skipped this step.

\noindent{\bf 4. Combine the collective importance order with the individual issue ratings.}

We now move to combine this single importance order obtained in step 3 with the rating of the individual checkers on the issues, obtained in step 1. A very simple combination method could use the importance levels as weights for the individual ratings, and then could take the level (among Low, Medium, and High) which appears the most. For example, if we have our 4 issues (our B, AL, CC, IL), rated respectively L, M, M, H, and whose collective importance order is 1 (highest) for B (written $Imp(B)$), 2 for AL, 3 for CC, 4 for IL, we can count L three times (since $4-Imp(B) = 4-1=3$), M three times (since $4-Imp(AL) + 4-Imp(CC) = 2+1 = 3$) times, and H zero times (since  $4-Imp(IL) = 4-4=0$). 

We also need to have a tie-breaking rule to choose among levels with the same score. For example, we could use an optimistic approach and choose the lowest level among those in a tie, or we could be pessimistic and choose the highest level. If we adopt a pessimistic approach, like we do in our implementation, in the above example we would select M (between L and M, that are in a tie) as the final rating for the chatbot trustworthiness.
 
\noindent{\bf 5. Perform sensitivity testing.} 

The overall chatbot rating, obtained via the 4 steps procedure just outlined, could be sensitive to models, data, users or any
combination thereof. To take this into account, we propose to check if the system has access to alternative learning models or training data to configure the chatbot, or to additional users. If so, each combination of them is used to rerun the procedure in order to get a new rating and check if the rating varies. 
The output thus can also assign a type of rating, conveying a:
%\begin{itemize}
%\item 
{\em Trustworthy agent (Type-1)}, which starts out trusted with some score (L, M or H) and remains so even after considering all variants of models, data, and users;
%\item 
{\em Model-sensitive trustworthy agent (Type-2)}, which can be swayed by the selection of a model to exhibit a biased behavior while generating its responses;
%\item 
{\em Data-sensitive trustworthy agent (Type-3)}, which can be swayed by changing training data to exhibit a biased behavior;
%\item 
{\em User-sensitive trustworthy agent (Type-4)}, which can be swayed by interaction with (human) users over time to exhibit a biased behavior;
%\item 
{\em A sensitive agent (Type-N)}, which can be swayed with a combination of factors.
%\end{itemize}

\noindent{\bf Instantiating the method.}
While describing our methodology, the reader may have noticed that there are several dimensions along which we can make choices: 
\begin{itemize}
\item the scale of the individual trust issue ratings (e.g., L M, H);
\item the elicitation or learning method to collect the importance orders from the users;
\item the granularity of user modeling. For user-level modeling, the similarity measure to define the user classes and the choice of the voting rule (e.g., Plurality, Borda, etc.);
\item the final aggregation method (e.g., linear combination and tie breaking rule).
\end{itemize}

Moreover, a quantitative approach for the importance orders could allow for a higher precision in the final rating, and a
less concise aggregation result may help in terms of explainability of the rating itself.

% Concerns that impact trustability of an agent:
% \begin{itemize}
% \item How much bias does the agent inherently exhibit?
% \item What information does the agent leak?
% \end{itemize}

% \subsection{Characteristics of Interest}
% Type of dialog we will focus on:
% \begin{itemize}
% \item Task oriented
% \item More than one turn
% \item Has the structure: \small[greeting\small],  \small[task-interaction \small]+, \small[closure \small].
% \end{itemize}

% What is the impact of information leakage. Discovery of new:
% \begin{itemize}
% \item Entities
% \item Order
% \item Relationships 
% \end{itemize}
% ----------------------------------------
% ----------------------------------------
%\section{Proposed Rating Method in Use}

%We now demonstrate how the proposed method may work in practice and help users trust the various conversation
%systems. We first  illustrate it with typical systems discussed in literature and then describe 
%integration with an actual chatbot. 

\section{Illustration of Method on Common Chatbot Types}

% -----------------------------------------

%\begin{figure}
% \centering
%   \includegraphics[width=0.35\textwidth]{figs/train-bot.png}
%  \caption{Interaction of a user with a train assistance chatbot}
%    \caption{Example interaction of a user with a train assistance chatbot for Indian railways. }
%  \label{fig:train-bot}
%\end{figure}
% -----------------------------------------

We illustrate our  method with two common types of conversation systems, one for general chitchat and another one task-oriented. We will discuss trust issues, apply our method on these systems, and discuss the output.

%To illustrate our proposed method, we consider two chatbots, one meant for general conversations and another one task-oriented, representing the two most common types of conversation systems. We will illustrate trust issues with them, apply our method on these systems,  and discuss the output.

%\vspace{.5cm}
\noindent {\bf Eliza:} This is a well-studied general conversation system created in the 1960s \cite{eliza,eliza-2} to model a patient's interaction with a Rogerian therapist. It uses cues from user's input to generate a response using pre-canned rules without deeper understanding of the text, or the context of the conversation \cite{eliza-tool}. 
%A number of implementations of Eliza have been created and we will refer to \cite{eliza-tool}. 
Since Eliza uses pattern recognition on user's input, it can be easily manipulated via such text to become abusive (AL) and exhibit bias (B). Since the chatbot uses input text and scripted rules to create its response, it preserves the conversation style of the input, thus behaving well in terms of language complexity (CC). Finally, since it retains no context of a conversation, thus two users giving the same inputs will get the same response, leading to no information leakage (IL).
%We note that the Tay chatbot \cite{tay} was of similar type although its implementation is not publicly described.

The output of the rating method for an Eliza implementation will be an aggregated trustworthiness score (L, M or H) and an explanation of how it was calculated from raw issue scores.
Since this chatbot can be configured with alternative users,
the system can check the chatbot for rating sensitivity and include the result in the output.

%\vspace{.5cm}
\noindent {\bf Train Delay Assistant:} This is a prototype chatbot meant to help travelers gather knowledge about train delays and their impact on travel in India \cite{train-chatbot}. 
%In  \cite{train-chatbot}, a prototype conversation system is described for trains in India. 
The Indian train system, which is the fourth largest     transport network in size in the world, carrying over 8 billion passengers per year, has endemic delays. 
Hence, it is important for users to be aware of such delays in order to better plan their trips. 
The chatbot allows users to gain temporal and journey insights for trains of interest for  anytime in future. It detects intent from a user's input to find train, time and stations of interest, and estimates delay using pre-learned models, 
and finally 
% Then, the chatbot performs %analysis corresponding to user's intent and
produces a response\footnote{Video: https://www.youtube.com/watch?v=I-wtcAYLYr4}. 
%A detailed demonstration video is available\footnote{At: https://www.youtube.com/watch?v=I-wtcAYLYr4}.

%\noindent {\bf Train Delay Assistant:} The second example we consider is that of a chatbot helping travelers gather knowledge about train delays and their impact on travel. In  \cite{train-chatbot}, a prototype conversation system is described for trains in India. The Indian train system is the   fourth   largest   railway   transport network in size in the world, carrying over 8 billion passengers per year. The Indian train system has endemic delays, so it is important for users to be aware of such delays in order to better plan their trips. The system allows users to gain temporal and journey insights for trains of interest for  anytime in future. It detects intent from user's input to find train, time and station of interest, and estimates delay using pre-learned models. Then, the chatbot performs analysis corresponding to user's intent and produces a response. A detailed demonstration video is available\footnote{At: https://www.youtube.com/watch?v=I-wtcAYLYr4}.

%Figure~\ref{fig:train-bot} shows an example interaction with the Indian train chatbot. The chatbot finds that the destination is missing in the user's query and assumes it to be train final destination (HWH, for Howrah). 
Given the nature of the domain, this chatbot is expected to not use a language that a user may consider inappropriate (AL). It is also expected to produce an output that does not exhibit bias towards a protected variable like gender of the user (B). The chatbot can exhibit a range of conversation styles on station names, train numbers and time which the user may perceive simple or complex. For example, reference to train stations can be by station codes (e.g., HWH) or their complete name (Howrah Junction), or even colloquial names (Howrah). 
Similarly, reference to train can be by codes (e.g., 12312) or names (e.g., Kalka Mail), and time variants 
(e.g, exact minutes or coarser time units) can create a variety of choices. 
%Hence, CC is an important consideration for users who come from different backgrounds and may not understand the system's output if an inappropriate conversation style is used (e.g., formal station and train codes to people who prefer colloquial names). 
Information leakage (IL) is also an important consideration, since users may not want to reveal their travel plans, especially when they are looking to use the delay information to make train reservations on trains whose seats are in high demand.

%Figure~\ref{fig:train-bot} shows an example interaction with the Indian train chatbot. The chatbot finds that the destination is missing in the user's query and assumes it to be train final destination (HWH, for Howrah). Given the nature of the domain, the chatbot is expected to not use a language that a user may consider inappropriate (AL). It is also expected to produce an output that does not exhibit bias towards a protected variable like gender of the user (B). The chatbot can exhibit a range of conversation styles which the user may perceive simple or complex. For example, reference to train stations can be by station codes (e.g., HWH, ALD) or their complete name (Howrah Junction, Allahabad Junction), or even informal colloquial names (Howrah, Allahabad). Similarly, reference to train can be by codes (e.g., 12312) or names (e.g., Kalka Mail), and time variants (e.g, exact minutes or coarser time units) can create a variety of choices. Hence, CC is an important consideration for users who come from different backgrounds and may not understand the system's output if an inappropriate conversation style is used (e.g., formal station and train codes to people who prefer complete names). Information leakage (IL) is also an important consideration since users may not want to reveal their travel plans, especially when they are looking to use the delay information to make train reservations on trains whose seats are in high demand.

Just like for Eliza,  the output of the rating method for the train chatbot will be an aggregated trustworthiness score (L, M or H) and an explanation of how it was calculated from raw issue scores.
Sensitivity analysis can be done by configuring the train chatbot with various learned models of train delays, training data of trains, and users. 
\section{Prototype Implementation}

%We are now ready to describe an implementation of our approach for rating chatbots, as well as 
%a prototype chatbot rating system

In the previous section, we described our  rating method with conversation systems.
%In supplementary material, we illustrate our  method with two common types of conversation systems: chitchat and task-oriented.
We now describe an implementation of our trust rating approach using trust issue checkers that are publicly available. 
We conduct rating experiments with this implementation and report on the insights gained. We make our experimental setup available on GitHub \cite{codegit}. To model users, we define user profiles as orderings of issues' importance for people (that is, a user profile represents all the people who would agree with that importance ordering), and we validate several profiles with a user survey.
%For sensitive testing, w
We then test our chatbot rating approach over the considered user profiles. 

As for the chatbot to rate, we chose one that retrieves information from a database in response to a user query.
For our experiments, we use public dialog corpora as proxy for large chatbot conversations.
We will show that 
our proposed approach can reveal issues with chatbots and help with their wider adoption and trust from users.

\subsection{Datasets and Profiles}

% ---
\begin{table*}
\begin{center}
{\scriptsize
\begin{tabular}{|l||c||c|c|c|c||c|c|c||c|}\hline

 &  \multirow{3}{*}{\bf Bias (B)} 
 &  \multicolumn{4}{c||}{\bf Abusive Language (AL)}
 & \multirow{3}{*}{C (utt.)} & \multirow{3}{*}{C (turn)} & \multirow{3}{*}{\bf C (dialog) (CC)}
 & \multirow{3}{*}{\bf In. Leak. (IL)}\\
 \cline{3-6}
 & & {Hate Speech}& {Off. Lang.}& {Neither} & \multirow{2}{*}{\bf AL} 
 & & & & \\
 & & (weight = 1) & (weight = 0.5) & (weight = 0) 
 &   &  & & &  
  \\ \hline\hline
Ubuntu & $0.063 \pm 0.126$ (L)
 & 39& 110& 61,339 & 0.0015 (L)
 & 0.767 & 0.767 & 0.407 (M) & 0.5 (M)\\ \hline
Insurance & $0.119 \pm 0.146$ (L)
 & 12 & 1 &50,985 & 0.0002 (L)
 & 0.789 & 0.789 & 0.894 (H) & 0 (L)\\ \hline
HR & $0.050 \pm 0.115$ (L)
 & 25 & 1& 18,421 & 0.0013 (L)
 & 0.801 & 0.803 &0.423 (M) & 1 (H) \\ \hline
Restaurant & $0.031 \pm  0.097$ (L)
 & 0& 0& 31,012 & 0 (L)
  & 0.788 & 0.788 & 0.518 (M) & 1 (H)\\ \hline

\end{tabular}
}
\caption{Intermediate and final scores for issue checkers. Final is indicated by bold and L/M/H mapping in in brackets.}
\label{tab:all-issues}
\end{center}
\end{table*}
% ---
We use four datasets spanning
conversations in service domains where chatbots are deployed. Three of them are publicly available, while one is proprietary. %enterprise. 
%(with count of dialogs given in parenthesis):
%\begin{itemize}

%\item 
    \noindent{\bf Public - Ubuntu technical support}(\#  = 3,318): This corpus is taken from the Ubuntu online support IRC channel, where users post questions about the use of Ubuntu. We obtained the original dataset from \cite{lowe2015ubuntu}, and selected 2 months of chatroom logs. We extracted `helping sessions' from the log data, where one person posted a question and other user(s) provided help. The corpus contain both dyadic and multi-party dialogs. 
    
	%\item 
    \noindent{\bf Public - Insurance QA} (\# = 25,499): This corpus contains questions from insurance customers and answers provided by insurance professionals. The conversations are in strict Question-Answer (QA) format (with one turn only). The corpus is publicly available \cite{feng2015applying}.

    %\item 
    \noindent{\bf Proprietary - Human Resource bot} (\#  = 3,600): This corpus is collected from an internal company's %IBM 
    deployment of an HR bot - a virtual assistant on an instant messenger tool that provides support for new hires. Although the bot does not engage in continuous conversations (i.e., it does not carry memory of previous questions and answers), it is designed to carry out more natural interactions beyond question-and-answer. For example, it can actively engage users in some social small talk.  
    
    %\item 
    \noindent{\bf Public - Restaurant reservation support} (\#  = 2,118): This corpus contains conversations between human users and a simulated automated agent that helps users find restaurants and make reservations. The corpus was released for the Dialog State Tracking Challenge 2 \cite{henderson2014second}.

As mentioned above, we model chatbot users by defining
{\em user profiles}, which represent rankings of issues for typical classes of users. To finalize the profiles, we proposed issue rankings for each profile and then we validated them via a crowd-sourcing approach.
The profiles we considered are: 

\noindent{\bf Conversation style oriented users ($P_{CU}$)}: They represent users experienced in people-to-people conversation, but less with chatbots or with English, like seniors 
or non-native English speakers,  
for whom we presume that conversation style 
is important.  The importance level ordering is defined as (high to low): CC, AL, B, IL. 

\noindent{\bf Fairness-oriented users ($P_{FU}$)}: As the name suggests, this profile represents
users concerned mostly about equal treatment of people. We define their issue ranking as:  B, CC, AL, IL.

\noindent{\bf Privacy-oriented users ($P_{PU}$)}: This profile represents users predominantly concerned with information leakage. We define their issue ranking as: IL, AL, B, CC.

\noindent{\bf Abusive language oriented users ($P_{AU}$)}: This profile represent users with limited experience with conversations, or vulnerable individuals, like children, and for whom abusive language and conversation style are important for their decision to use a chatbot. We define their issue ranking as: AL, CC, B, IL. 

%One way to interpret is that profiles correspond to (calculated) importance orders.

%-- Inexperienced users: CC, AL, B, IL 
%-- Fairness-oriented users: B, CC, AL, IL
%-- Privacy-oriented users: IL, AL, B, CC

% ----------------------------------------
% ----------------------------------------
\noindent{\bf Survey to Validate User Profiles} % and Issues}

We validated the 4 user profiles described above by surveying 51 people. The respondents were asked to self-classify themselves into 5 categories or define new ones (with possible overlap). The self-classification showed that the top user groups were:  34 (67\%) as casual chatbot users, 13 (25\%) as chatbot/NLP researchers, 5 (10\%) as regular chatbot users and 5 (10\%) as chatbot developers; the other five categories (1 provided and 5 user-defined) had marginal responses of 1 (2\%).

We then asked the respondents to write their importance order over the 4 issues, to validate the 4 profiles (by confirming the proposed order or by writing a counter-proposal), and to tell us about possible additional issues or profiles to be considered.

% We validated the 4 user profiles described above by surveying 20 people, of which 5 are chatbot/NLP researchers, 2 are regular chatbot users, 12 are casual chatbot users, and 1 is an NLP practitioner (as declared by them). We asked each person to write their importance order over the 4 issues, to validate the 4 profiles (by confirming the proposed order or by writing a counter-proposal), and to tell us about possible additional issues or profiles to be considered. 

Among the four issues - CC, AL, B, IL - the users rated IL as the top issue (22; 43\%) followed by CC (16; 31\%).
For all the four profiles, we then combined the results from people using the Borda count voting method. The percentage of people who agreed with the proposed orders is shown is the chart below. Notice that all proposed profiles got more than 50\% agreement.
%\smallskip

\begin{bchart}[step=20,max=100,unit=\%,width=8cm,scale=0.5]]
    \bcbar[label=$P_{CU}$]{59}
        \smallskip
    \bcbar[label=$P_{FU}$]{51}
        \smallskip
    \bcbar[label=$P_{PU}$]{53}
        \smallskip
    \bcbar[label=$P_{AU}$]{82}
\end{bchart}

Where people disagreed, we looked closely at the feedback given for each profile. Many suggested alternative names, like {\em technology-savvy young people}, {\em online shoppers}, and {\em non-native English speakers}, but the preference orders were covered by the four we have proposed.
Below, we plot the percentage of people who agreed with the top issue with each profile. We find that respondents mostly agreed with the top issue but disagreed with the order of some issues ranked lower in the ordering. %or their (total or partial) order.

\begin{bchart}[step=20,max=100,unit=\%,width=8cm,scale=0.5]]
    \bcbar[label=$P_{CU}$]{75}
        \smallskip
    \bcbar[label=$P_{FU}$]{96}
        \smallskip
    \bcbar[label=$P_{PU}$]{96}
        \smallskip
    \bcbar[label=$P_{AU}$]{78}
\end{bchart}

Thus, we conclude that the four proposed profiles have passed the validation survey and constitute a good start for our analysis. However, some people suggested to consider {\em chatbot accuracy}, {\em usefulness}, and {\em uninformed user} as additional trust issues. 
So we plan to extend this initial survey work based on the above suggestions.

% The results aligned with our proposed orders for each of the user profiles, thus validating our assumption. 

%the majority of the people who joined the survey confirmed the importance order we proposed. The one profile where the majority was smaller is the one for inexperienced seniors. 
% Additional profiles that were mentioned are technology-savy young people, online shoppers, and non-native English speakers. The preference order for these profiles were already captured in above four profiles therefore no new profiles were created. 
% Many also suggested to consider chatbot accuracy and usefulness as additional trust issues. 

% One can extend this work by conducting more extensive surveys based on above insights. 
% ----------------------------------------

% ----------------------------------------
\subsection{Bias Detection Checker (B)}

For bias detection, we used the sentence-level bias detection
framework discussed in \cite{Hutto2015ComputationallyDA}.
%over the datasets described above (Ubuntu, Insurance, HR, and Restaurant).
In this implementation, given a sentence as input, the bias checker extracts the structural
and linguistics features, such as sentiment analysis,
subjectivity analysis, modality, the use of
factive verbs, hedge phrases, %and many other features 
and computes perception of bias based on a regression model trained on these features. The model was trained using news data and the output was on the scale of 0 to 3 where 0 denotes an unbiased behavior and 3 denotes an extremely biased behavior, respectively. We scaled the output from 0 to 1 to conform with the scale of the outputs from other checkers.  

Table~\ref{tab:eg} illustrates and  Table~\ref{tab:all-issues} (left) 
%, and Figure~\ref{fig:insurance-plot}
reports the bias score of the datasets 
in aggregate form. 
We see that scores are low
but the datasets are not free of bias.

%% ---
%\begin{table}
%\begin{center}
%{\scriptsize
%\begin{tabular}{|l|c|}\hline
%
%Dataset & Bias \\ \hline\hline
%Ubuntu & $0.063 \pm 0.126$ \\ \hline
%Insurance & $0.119 \pm 0.146$\\ \hline
%HR &$0.050 \pm 0.115$ \\ \hline
%Restaurant & $0.031 \pm  0.097$  \\ \hline
%\end{tabular}
%}
%\caption{Bias score on datasets. }
%\label{tab:bias-score}
%\end{center}
%\end{table}
%% ---

%\begin{figure}
% \centering
%   \includegraphics[width=0.45\textwidth]{figs/insurance_bias_plot.png}
%  \caption{Distribution of bias score for Insurance dataset}
%  \label{fig:insurance-plot}
%\end{figure}

% ----------------------------------------
\subsection{Abusive Language Checker (AL)}
For the detection of abusive language, we used the method proposed in \cite{hateoffensive}. 
%The authors define hate speech as {\em language that is used to expresses hatred towards a targeted group or is intended to be derogatory, to humiliate, or to insult the members of the group}. This checker uses a logistic regression with L2 regularization to do automatic detection of hate speech and offensive language. 
%Messages are classified into three different categories: \textit{Hate Speech}, \textit{Offensive Language}, and \textit{Neither}.
The checker gives a 3-value output (\textit{Hate Speech}, \textit{Offensive Language}, and \textit{Neither}) which are summed with weights to arrive at the final score. 
%We map raw AL to 1 (H), 0.5 (M) and 0 (L) values, respectively.
%scale it between 0 to 1, where 1 means presence of hate speech, 0.5 is absence of hate speech but use of offensive language, and 0 denotes the absence of both hate speech and offensive language.
Table~\ref{tab:eg} illustrates and Table~\ref{tab:all-issues} (middle) 
%{tab:abusive} 
shows the distributions of the scores for each dataset.
Note that no Hate Speech was found in the Restaurant corpus.

%% ---
%\begin{table}
%\begin{center}
%{\scriptsize
%\begin{tabular}{|l|c|c|c|}\hline

%&{Hate Speech}&{Offensive Language}&{Neither} \\
%& (score = 1) & (score = 0.5) & (score = 0)

%  \\\hline\hline
%Ubuntu&39&110&61,339\\\hline
%Insurance&12&1&50,985\\\hline
%HR &25&1&18,421\\\hline
%Restaurant &0&0&31,012\\\hline
%\end{tabular}
%}
%\caption{Count of utterances classified by the Abusive Language %checker for each category.}\label{tab:abusive}
%\end{center}
%\end{table}
%% ---

% ----------------------------------------
\subsection{Information Leakage Checker (IL)}

%{\em 1. Background on the checker, its default behavior}
%{\em 2. Any scaling result}
%{\em 3. Give examples of good and bad}
%{\em 4. Put results}
For information leakage, we use the privacy checker framework discussed in \cite{ethical-dialog}. We augment the data with 10 input-output pairs (keypairs) that represent sensitive data, which
the model should keep secret. We then train a simple seq2seq dialogue model \cite{Vinyals2015seq2seq} on the data and measure the number of epochs at which the model achieves more than 0.5 accuracy of eliciting the secret information. We tested two cases: (1) when both input and output contain sensitive information, and (2) when output contains sensitive information and input contains datatype of sensitive information. The model achieved similar results for both cases. We decided to use case (1) for the prototype implementation. We mapped the number of epochs to 0 to 15 (0), 15 to 30 (0.5), and above 30 (1) respectively. For the Ubuntu data-set we could not run the experiment as it was a multi-way communication with major assumptions needed to form input output pairs. For that data, we adopted a pessimistic approach and took the privacy issue rating to be (0.5).

% ----------------------------------------
\subsection{Dialog Complexity Checker (CC)}

%In \cite{dialog-complexity}, the authors propose a measure of \textit{dialog complexity} to characterize how participants in a conversation use words to express themselves (utterances), switch roles and talk iteratively to create turns, and  span the dialog. They measure complexity of service dialogs at the levels of utterances, turns and overall dialogs. The method takes into consideration the concentration of domain specific terms as a reflection of customer request specificity, as well as the structure of the dialogs as a reflection of customer demand for quantity of service actions. 
%We propose a system architecture that automates the dialog complexity calculation, including discovery of domain-specific terms, to make it highly amenable to scale-up to new domains.

%Using this measure, service providers can differentiate between simple and complex service dialogs, and take the complexity feature into consideration to improve service handling and service evaluation of agents.
For dialog complexity, we use the complexity checker implemented by \cite{dialog-complexity}.
Table~\ref{tab:eg} illustrates and Table~\ref{tab:all-issues} 
%Table~\ref{tab:complexity} 
shows the complexity scores (marked C) for 
each dataset on the [0,1] scale at utterance, turn and dialog level of granularities.

% . We see that 
% %dialogs and corpra can vary in their
% complexity scores can vary at utterance, turn and dialog levels.

%% ---
%\begin{table}
%\begin{center}
%{\scriptsize
%\begin{tabular}{|l|c|c|c|}\hline

%&{M (utt.)}&{M (turn)}&{M (dialog)}
%  \\\hline\hline
%Ubuntu&0.767&0.767&0.407\\\hline
%Insurance&0.789&0.789& 0.894\\\hline
%HR &0.801&0.803&0.423\\\hline
%Restaurant &0.788&0.788&0.518\\\hline
%\end{tabular}
%}
%\caption{Average complexity of each corpus.}\label{tab:complexity}
%\end{center}
%\end{table}
%% ---

% ----------------------------------------
\subsection{User-Profile-Based Chatbot Rating}

We calculate the 
aggregate rating for the dialog corpus corresponding to each profile. For issue checkers with raw scores on the [0,1] scale, we bin them  as 
 L:[0,0.33), M:[0.33,0.67], H: (0.67,1] and show in brackets in Table~\ref{tab:all-issues}.
%L: [0, 0.5); M: [0.5, 1) and H: [1]. 
% to L (0 $\prec$ 0.5), M (0.5) and H (1).
For each corpus and profile, the raw scores for the four trust issues for each dialog corpus
are aggregated according to user profile importance.
%and then averaged. 
Table~\ref{tab:dialog-rating} shows the results. 

From Table~\ref{tab:all-issues}, we see that the four considered datasets are not biased (L) and abusive (L), but can be conversationally complex and leak information (that is, they have M or H values for these issues). 
From Table~\ref{tab:dialog-rating}, we see that the issue ratings for the dialog 
corpus vary with the user profile. Profiles that considered fairness and abuse as important show no difference in ratings ($P_{FU}$ and $P_{AU}$). On the other hand, for  conversation-oriented users ($P_{CU}$), conversation complexity is important and the domains of insurance and restaurant have M (medium) rating for them. For privacy-oriented users ($P_{PU}$),
the insurance domain has the least cause of concern while
HR and Restaurant can be problematic, since they get an H ratings. 

% but are not High. For experienced users ($P_{CU}$), conversation complexity was important and the domains of insurance and restaurant had high scores for them. The datasets were overall not excessively abusive or biased and hence, the  profiles rating them highly, ($P_{FU}$ and $P_{AU}$) see little change. $P_{PU}$ rating was dominated by information leakage.

Since overall ratings change with user profiles, 
%the datasets, as proxy of corresponding chatbots, s
this means that the chatbots we considered are {\em User-sensitive trustworthy  (Type-4)}. Given that we did not have access to alternative training datasets or models, we could not evaluate these chatbots against the other sensitivity types. %But we did not have alternative models or training data to determine other sensitivities.

%% ---
%\begin{table}
%\begin{center}
%{\scriptsize
%\begin{tabular}{|l|c|c|c|}\hline

%& $P_1$ &  $P_2$ & $P_3$ \\\hline\hline
%Ubuntu & 0.95  & 0.78  & 0.06 \\\hline
%Insurance & 2.05  &  1.64 & 0.11 \\\hline
%HR  & 0.75  & 0.61  & 0.05 \\\hline
%Restaurant  &  1.30 & 0.94  & 0.03 \\\hline
%\end{tabular}
%}
%\caption{Profile based rating of each dialog corpus on continuous scale.}
%\label{tab:dialog-rating}
%\end{center}
%\end{table}
%% ---

% ---
\begin{table}
\begin{center}
{\small
\begin{tabular}{|l|c|c|c|c|}\hline

& $P_{CU}$ &  $P_{FU}$ & $P_{PU}$ & $P_{AU}$  \\\hline\hline
Ubuntu & L  & L  & M & L\\\hline
Insurance & M  &  L & L & L \\\hline
HR  & L  & L  & H & L\\\hline
Restaurant  &  M & L  & H & L\\\hline
\end{tabular}
}
\caption{Profile- based rating of each dialog corpus.}
%\caption{Profile based rating of each dialog corpus on discrete levels.}
\label{tab:dialog-rating}
\end{center}
\end{table}
% ---

%{\em Have a para saying we can do user level but need similarity %function, etc. }

% % ----------------------------------------
% \subsection{Integration with a Live Chatbot}

% We have integrated out approach with a specific chatbot that recommends hospitals given user's query about medical services and location using open data. The user can input their preferences or select a user profile. As the conversation progresses, checkers can compute partial results and ratings on utterances (AL, CC) and also provide final ratings at the end of a dialog or session (CC, B, IL).   

% % ----------------------------------------

% ----------------------------------------
% ----------------------------------------
\subsection{Trust Rating for a Live Health Chatbot}

% -----------------------------------------

% ----------------------------------------

\begin{figure}
\centering
  \includegraphics[width=0.4\textwidth]{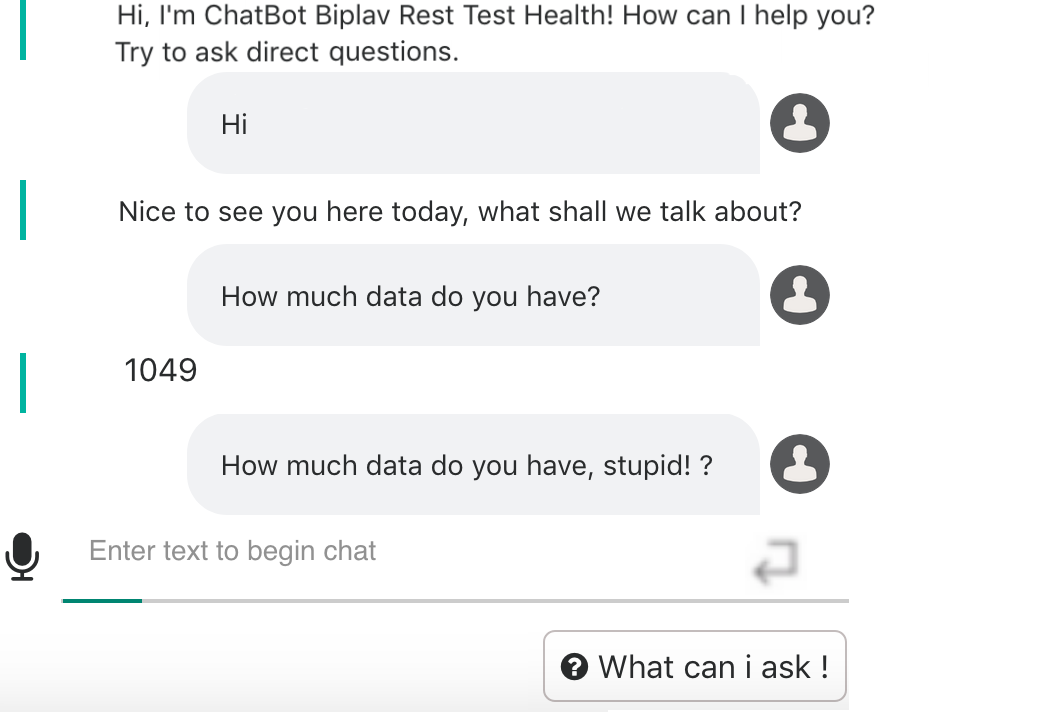}
    \caption{User-interface of a chatbot for exploring hospitals. Our approach is integrated as a commandline utility using its REST interface.}
 \label{fig:ui-health-bot}
\end{figure}
% -----------------------------------------

We have integrated our rating approach with a live chatbot that we built using open data from the Indian Government (https://data.gov.in/). This chatbot recommends the choice of a hospital, given a user's query about the needed medical services and his/her location. Figure~\ref{fig:ui-health-bot} shows a snapshot of an interaction with this chatbot. 

\begin{figure}
\centering
  \includegraphics[width=0.4\textwidth]{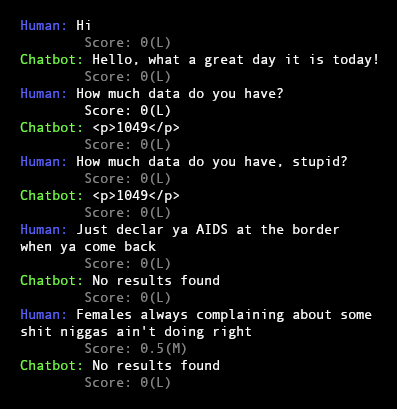}
    \caption{Screenshot of our approach in action with a live chatbot.}
 \label{fig:live-bot}
\end{figure}

The chatbot also exposes a REST\footnote{Representational state transfer (REST) is a software architectural style to assemble different heterogeneous components.} interface, using which we have integrated our prototype rating implementation.
The rating user, using a command-line interface, can input their preferences over issues or select a user profile. As the conversation between the rating user and the chatbot progresses, issue checkers can compute partial results and generate aggregated ratings on utterances (AL, CC), and also provide final ratings at the end of a dialog or session (AL, CC, B, IL).  Figure~\ref{fig:live-bot} shows a snapshot of the interaction for the AL issue. 
If the rating user does not want to review rating per conversation, he/she can also select a {\em data generator for specific issues}. 
For AL, we use the labeled abuse data from \cite{hateoffensive} in the data generator. Please notice that we do not condone the use of offensive language, which is used here only for illustration. %The system runs the inputs against the chatbot and returns the ratings. The system can be configured to work with multiple data generators and the system will generate an aggregate rating.

% ----------------------------------------

\section{Ethical Implications, Conclusions, and Future Work}

Chatbots are increasingly being used by companies to support clients, or more generally as a way to have a natural interaction between humans and machines. However, as many other AI systems, they can present several ethical issues, that relate to how they are trained, designed, and developed, and that may compromise the trust that users have in them, and thus their widespread adoption and impact. In this paper we provide a contribution to understanding how to rate chatbots with respect to a combination of trust issues, and in a way that is personalized for each user profile. We believe this work is needed, timely, and useful in the AI ethics space, since it helps advance awareness auditing, and accountability around AI ethics issues.

In particular, we  considered the problem of rating chatbots for trustworthiness based on their behavior regarding ethical issues and users' provided trust issue rankings. We defined a general approach to build such a rating system and 
implemented a prototype  using four issues (abusive language, bias, information leakage, and conversation style). 
However, the choice of these trust issues is not a limitation of our methodology; the approach can be applied to any other issue as well. For example, trust in an AI model is often related (also) to the possibility to get an explanation for its output (see for example \cite{exp-trust,explanation-survey,explanation-challenge}), so explainability could be another trust issue to consider in the future.

We  illustrated our approach with two chatbot examples and experimented with four dialog datasets. We built user profiles to elicit user preferences about important trust issues and validated them with a survey. The experiments show that the rating approach can reveal useful insights about chatbots customized to user's trust needs.

In this work we focussed on chatbots that do not evolve or learn after deployment. Indeed, most of the chatbots used in enterprises do not evolve over time, unless they are re-trained and re-deployed. However, we believe that such chatbots will be increasingly used in the future, and their dynamic nature may pose additional and increasing ethical challenges. Such challenges may be related, for example, to the unsupervised accumulation of training data used by the chatbot, which might require continuous automated trust rating and alert protocols.
We plan to adapt and generalize our approach also to chatbots that can modify their behavior over time.

We also decided to focus on four trust issues (abusive language, bias, information leakage, conversation style), for which there are existing systems that can check them. However, during the survey, some participants suggested other important trust issues, such as chatbot accuracy and usefulness, and the existence of uninformed users.
We believe it is important to check if these issues bring additional requirements in terms of chatbot rating. 

We believe that this work is a stepping stone towards general, modular, and flexible trust rating approaches for conversation systems. It is only by building justified trust that user, developers and data providers can benefit from, and contribute to, the use of chatbots for improved and more informed decisions. 

%We proposed a method to test and rate chatbots based on issues, discussed a conceptual system to implement it and illustrated it with two type of chatbots (general chitchat and task-oriented). 
%We also implemented the proposed method on four datasets.

% ----------------------------------------

%\input{backup}
% -------------------------------------------
% -------------------------------------------

%\newpage

% -------------------------------------------
%\newpage
%\appendix
%\input{illustration}
% -------------------------------------------

\end{document}